\documentclass[oneside,final,onecolumn,titlepage,thmsa,letterpaper]{article}
\usepackage{sw20lart}

\input tcilatex
\QQQ{Language}{
American English
}

\begin{document}

\begin{center}
\textbf{QuasiCerenkov Radiation of Relativistic Electrons in Crystals in the
Presence of External Excitations}
\end{center}

\begin{quotation}
H.A.Aslanyan$\footnote{%
E-mail: aharut@iapp.sci.am, Tel/Fax: (3742) 24 58 85}$, A.R.Mkrtchyan,
A.H.Mkrtchian

Institute of Applied Problems of Physics of Armenian NAS,

25, Hr.Nersesian St., 375014, Yerevan, Armenia
\end{quotation}

\begin{quote}
Abstract

The paper is devoted to the study of the influence of crystalline lattice
distortions due to external excitations (acoustic vibrations, temperature
gradient, etc.) on the Quasicerenkov radiation. Equations describing
Quasicerenkov radiation of charged particles in distorted crystals are
derived. These equations are solved numerically. It is shown that certain
types of lattice deformations may intensify the Quasicerenkov radiation by
several times.
\end{quote}

It is known that being in uniform linear motion a charged particle radiates
only if the Cerenkov condition is satisfied or the medium has space and/or
time inhomogeneities. In the case of space inhomogeneities intensity of the
radiated beam, its direction, and frequency are depended on the type and
size of inhomogeneities. If these inhomogeneities are periodically arranged
then the radiation, with the wavelength of the range of this periodicity,
may be intensified due to the interference of the waves radiated from the
different inhomogeneities. Such radiation takes place when a charged
particle moves through a crystal. Radiation emitted under the Bragg angles
is formed by diffraction of the secondary waves accompanying the charged
particles on the crystalline lattice. It is called Quasicerenkov radiation 
\cite{gar2} (QCR) (or Parametric x-ray radiation \cite{bar}). The QCR was
predicted \cite{gar},\cite{bar} in 1971 and their properties have been
studied on the basis of semi classical arguments in many theoretical works
(for example see \cite{gar2}). The experimental observation of QCR occurred
after 1985 (see Refs \cite{vor}-\cite{mkr1}). The results of this and
following experiments in the main agree with the theory. Recent works (see 
\cite{fior}) have revealed some disagreement of the measured and theoretical
values of the ratios of higher order radiation intensities to first order
radiation intensity for mosaic graphite.

The existing QCR theory was developed only for perfect (dynamic theory) and
mosaic (kinematic theory) crystals that can not describe the dynamic effects
in the presence of week distortions in the crystal. In the present paper an
attempt is made to develop a new method for investigating the influence of
lattice distortions on QCR.

The QCR phenomenon is described by Maxwell's equations, where the
permittivity of the medium is considered to be a periodic function of the
spatial coordinates. The Fourier-transform of the electromagnetic induction
with respect to the time is found to be of the form $\vec D=\vec D_s+\vec
E_e $, where $\vec E_e$ is the field of the moving charge in vacuum, and $%
\vec D_s$ is the scattered field. Then the Maxwell equations are reduced to

\begin{equation}
\Delta \vec D_s+\frac{\omega ^2}{c^2}\vec D_s+rotrot(\chi \vec
D_s)=-rotrot(\chi \vec E_e),  \label{max}
\end{equation}
where $\chi (\vec r,\omega )$ is the polarizability of the medium, $\omega $
is the radiation frequency and for $\vec E_e$ it is easy to find

\begin{equation}
\vec E_e=\frac{ie}{2\pi ^2v}\iint \frac{\gamma ^{-2}\vec k+\vec q}{\gamma
^{-2}k^2+q^2}exp(-i\vec k\vec r-i\vec q\vec r)d\vec q,  \label{Ee}
\end{equation}
where $e$ is the particle charge, $\gamma $ is the Lorentz factor, $\vec
k=\omega \vec v/v^2$, $\vec v$ is the particle velocity, while integration
is over the vectors perpendicular to the particle path ($\vec q\perp \vec v$%
). As it is seen from (\ref{Ee}), the field of the moving charged particle
can be interpreted as a sum of secondary waves with wave vectors $\vec
k+\vec q$. When the crystal is distorted in a way that the characteristic
length of the lattice deformations exceed many times the sizes of elementary
cell then for $\chi (\vec r,\omega )$ we can write \cite{tag}

\begin{equation}
\chi (\vec r,\omega )=\sum_b\chi _bexp(-i\vec b(\vec r-\vec u)),  \label{pol}
\end{equation}
where summation is performed with respect to the reciprocal lattice vectors $%
\vec b$, $\vec u$ is the vector displacement of the elementary cell from its
initial position. Taking into account (\ref{pol}) it is convenient to search
a solution of (\ref{max}) in the form

\begin{equation}
\vec D_s=\frac{ie}{2\pi ^2v}\sum_g\iint \vec D_gexp(-i\vec k_g\vec r+i\vec
g\vec u)d\vec q,  \label{eq4}
\end{equation}
where $\vec k_g=\vec k+\vec q+\vec g$ and $\vec D_g$ are slow functions of
the coordinates. Substituting (\ref{eq4}) into (\ref{max}) and using the
principle of superposition, after multiplying both sides of the equation by $%
exp(i\vec k_h\vec r+i\vec h\vec u)$ and integrating over the elementary cell
volume we obtain an infinite system of equations

\begin{equation}
2(\vec k_h\vec \nabla )\vec D_h+i(\frac{\omega ^2}{c^2}-k_h^2(1-\chi
_0)+2(\vec k_h\vec \nabla )(\vec h\vec u))\vec D_h-i\sum_{b\neq h}\chi
_{h-b}\vec k_h\times \vec k_h\times \vec D_b=i\vec F_h,  \label{eq5}
\end{equation}
where 

\[
\vec F_h=\frac{\chi _h}{\gamma ^{-2}k^2+q^2}[\vec k_h,[\vec k_h,\gamma
^{-2}\vec k+\vec q]],
\]
and $\vec h$ takes all the values of the reciprocal lattice vectors. When
deriving these equations we neglect small terms of second and higher order
(i.e., the terms containing second derivatives of the slow functions $\vec u$
and $\vec D_h$, the production of first derivatives by one another, and by $%
\chi _h$ as for x-ray frequency range $|\chi |\symbol{126}10^{-6}$). The
left-hand side of system (\ref{eq5}) completely coincides with the Takagi
equations \cite{tag} for x-ray diffraction in distorted crystals. Let us
consider the case, when for a given frequency only two strong waves
scattered in the directions $\vec k_0$ and $\vec k_h$ exist. In this case
only two equations remain in system (\ref{eq5}) describing the amplitudes $%
\vec D_0$ and $\vec D_h$. After separating the radiation into the normal and
coplanar polarizations one can find the separate systems of two equations
for every kind of polarization

\begin{equation}
\begin{tabular}{c}
$2(\vec k_0\vec \nabla )D_0^\alpha +ia_{00}D_0^\alpha +ia_{0h}^\alpha
D_h^\alpha =iF_0^\alpha ,$ \\ 
$2(\vec k_h\vec \nabla )D_h^\alpha +i(a_{hh}-2(\vec k_h\vec \nabla )(\vec
h\vec u))D_h^\alpha +ia_{h0}^\alpha D_0^\alpha =iF_h^\alpha ,$%
\end{tabular}
\label{eq6}
\end{equation}

where index $\alpha =\sigma ,\pi $ indicates the polarization type ($\sigma $
corresponds to the normal polarization when the amplitudes are perpendicular
to the plane composed by wave vectors $\vec k_0$ and $\vec k_h$, and $\pi $
corresponds to the coplanar polarization when they are in that plane),

\[
\begin{tabular}{cc}
$a_{00}=\chi _0k_0^2-(\gamma ^{-2}k^2+q^2),$ & $a_{hh}=\chi _0k_h^2-(\gamma
^{-2}k^2+q^2)+k_0^2-k_h^2,$%
\end{tabular}
\]

\[
\begin{tabular}{cccc}
$a_{0h}^\sigma =k_0^2\chi _h,$ & $a_{0h}^\pi =k_0^2\chi _h\cos 2\theta ,$ & $%
a_{h0}^\sigma =k_h^2\chi _h,$ & $a_{h0}^\pi =k_h^2\chi _h\cos 2\theta ,$%
\end{tabular}
\]

$\cos 2\theta =(\vec k_0\vec k_h)/(k_0k_h)$ and the vector amplitudes $\vec
D_{0,h}$, $\vec F_{0,h}$ are defined by the scalar amplitudes $%
D_{0,h}^\alpha $, $F_{0,h}^\alpha $ by the expression

\[
\vec A_{0,h}=(A_{0,h}^\sigma [\vec k_0,\vec k_h]+A_{0,h}^\pi [\vec
k_{0,h},[\vec k_0,\vec k_h]]/k_{0,h})/(k_0k_h\sin 2\theta ), 
\]
where $\vec A=\vec D$ or $\vec F$. To solve the problem of finding a
relativistic electron's QCR field it is necessary to specify boundary
conditions for the system (\ref{eq6}). For the Laue case of orientation the
boundary conditions for the two wave approximation is 
\begin{equation}
\vec D_0(\vec r_p)=\vec D_h(\vec r_p)=0,  \label{eq7}
\end{equation}
where $\vec r_p$ is the radius vector of a point on the crystal entrance
surface, as is no radiation field before the crystal. The number of $\gamma $%
-quanta with the energy $\hbar \omega $ emitted in the direction $\vec k_h$
is

\begin{equation}
\frac{\partial N_h}{\partial \omega }=\frac{c\cos \theta }{4\pi \hbar \omega 
}\iint (|D_h^\sigma |^2+|D_h^\pi |^2)dxdy,  \label{eq8}
\end{equation}
where $(x,y)$ are the coordinates of exit surface of the crystal. Equations (%
\ref{eq6}) with the boundary conditions (\ref{eq7}) may be analytically
solved only for certain types of distortions of crystals. It should be
mentioned that since these equations without right-hand side are the same as
for the case of x-ray diffraction in the distorted crystal \cite{tag} and
since the solution of inhomogeneous equations can be built by the solutions
of homogeneous part of that equations then the problem of QCR is
analytically solvable for each type of distortions for which the problem of
x-ray diffraction is analytically solvable. For example they may be solved
analytically in the case of quadratic deformations of crystalline lattice 
\cite{Gab} (that is, in the case of temperature gradient or crystal
bending). In common case they will be solved approximately by analytic or
numeric methods. We have studied obtained QCR equations by numeric methods
for two practically interesting cases of the crystal's distortions described
in \cite{mkr3}. In the first case

\begin{equation}
\vec h\vec u=\frac{2\pi u_0}d\sin (\frac{\pi z}T).  \label{eq9}
\end{equation}

These type of distortions are generated when a piezocrystal is excited by an
alternating voltage with the resonant frequency of the sample (the time
dependence in (\ref{eq9}) is omitted, as the time of particle transmission
trough the crystal is much less than the period of the acoustic vibrations).
In the second case the crystal is heated on one side and is cooled on the
other one, so that the direction of the temperature gradient is
perpendicular to the reflection planes, and the function $\vec h\vec u$ has
the form

\begin{equation}
\vec h\vec u=\frac{2\pi u_0}d\frac{4\pi z}T(1-\frac zT).  \label{eq10}
\end{equation}

In both cases the crystal is oriented by the symmetric Laue geometry when
the diffraction vector is parallel to the entrance surface of crystal. In
these cases $\vec h\vec u$ depends only on the coordinate perpendicular to
the diffraction vector $\vec h$, and the equations (\ref{eq6}) can be
reduced to the system of ordinary differential equations. The calculations
are carried out by the Runge-Kutta numeric method for the parameter values
according to experimental data of \cite{mkr3}. The results for acoustic
vibration case are presented in the Figs. Fig.1 shows the energy or the
frequency dependence of the number of radiated $\gamma $-quanta emitted in
the diffraction direction for various values of the amplitude of the
acoustic vibrations. Fig. 2 shows the dependence of the integral number of
the emitted QCR $\gamma $-quanta on the amplitude of the acoustic vibrations.

\FRAME{itbpFU}{2.1594in}{2.1594in}{0in}{\Qcb{Fig 1. The frequency dependence
of QCR radiated photons' number for different values of acoustic vibrations'
amplitude ($\nu =0$ is equivalent to $E\gamma =10.1KeV$): a)$u_0/d=0$; b) $%
u_0/d=30$; c) $u_0/d=60$ ($d$ is the interplane distance).}}{\Qlb{Fig 1}}{%
xxxfig1.bmp}{\special{language "Scientific Word";type
"GRAPHIC";maintain-aspect-ratio TRUE;display "USEDEF";valid_file "F";width
2.1594in;height 2.1594in;depth 0in;cropleft "0";croptop "0.9983";cropright
"0.9983";cropbottom "0";filename
'C:/HARUT/WORK/MYDOC/PAPERS/XXXFIG1.BMP';file-properties "XNPEU";}}

\FRAME{ihFU}{2.1594in}{2.1594in}{0in}{\Qcb{Fig 2. The integral number of
diffracted $\gamma $-quanta depended on the acoustic vibration's amplitude.}%
}{\Qlb{Fig 2}}{xxxfig2.bmp}{\special{language "Scientific Word";type
"GRAPHIC";maintain-aspect-ratio TRUE;display "USEDEF";valid_file "F";width
2.1594in;height 2.1594in;depth 0in;cropleft "0";croptop "0.9732";cropright
"0.9742";cropbottom "0";filename
'C:/HARUT/WORK/MYDOC/PAPERS/XXXFIG2.BMP';file-properties "XNPEU";}}

As it is seen from Figs, the QCR intensity increases several times with the
increase of the amplitude of the acoustic vibrations. For the high values of
the vibration amplitude the intensity curve goes to the saturation. The
similar results are obtained in the case of temperature gradient. These
results are in good agreement with the experimental results of \cite{mkr3}
and \cite{avag}.

\end{document}